\documentclass{PoS}
\usepackage{graphicx}
\usepackage[mathscr]{euscript}
\usepackage{amsmath}
\usepackage{amsfonts}
\usepackage{physics}
\usepackage{latexsym,amsmath,amstext}
\usepackage{mathrsfs}
\usepackage[utf8x]{inputenc}
\usepackage{epsfig}
\usepackage{float}
\usepackage{slashed}
\usepackage{amssymb,fontenc,times,mathptmx,graphicx}

\title{Pion structure from Lattice QCD}

\ShortTitle{Pion structure}

\author{
  \speaker{Peter Petreczky}$^{1}$, Taku Izubuchi$^{1,2}$, Luchang Jin$^{2,3}$,
  Christos Kallidonis$^{4}$, Nikhil Karthik$^{1}$, Swagato Mukherjee$^{1}$,
  Charles Shugert$^{1,4}$, Sergey Syritsyn$^{2,4}$\\
\llap{$^1$}Physics Department, Brookhaven National Laboratory, Upton, NY 11973, USA \\
\llap{$^2$}RIKEN-BNL Research Center, Brookhaven National Lab, Upton, NY, 11973, USA \\
\llap{$^3$}Physics Department, University of Connecticut, Storrs, Connecticut 06269-3046, USA \\
\llap{$^4$}Department of Physics and Astronomy, Stony Brook University, Stony Brook, NY 11794, USA
}

\abstract{We present preliminary study of parton distribution
inside the pion using mixed action approach with HYP smeared
valence clover quarks on HISQ sea within the framework of Large Momentum Effective Theory. 
We use 2+1 flavor $48^3 \times 64$ HISQ
lattices with lattices spacing of $a=0.06$ fm and valence quark
masses corresponding to pion mass of 300 MeV.
          }

\FullConference{XIII Quark Confinement and the Hadron Spectrum - Confinement2018\\
		31 July - 6 August 2018\\
		Maynooth University, Ireland}

\begin{document}

\section{Introduction}
QCD factorization implies that the cross-section of hard hadronic processes can
be written in terms of convolution of  partonic cross-section and parton distribution 
functions. In the case of quarks the parton distribution function (PDF) 
can be defined in terms of matrix
element of quark bilinear operator of the fast moving hadron
\begin{equation}
  q(\xi) = \frac{1}{4\pi}\int d\xi^{-} e^{ixP^{+}\xi^{-}}
  \bra{H(P)}\bar\psi(\xi^-)\gamma^+W_L(\xi^-, 0)\psi(0)\ket{H(P)},~P \rightarrow \infty
\end{equation}
where
$W_L(\xi^-, 0) = e^{ig\int_0^{\xi^-} d\xi^- A^+}$ is a straight Wilson Line on the
light-cone, and $\xi^{\pm} = (t \pm z)/\sqrt{2}$. 
First principle calculation of PDF is not possible because lattice QCD is formulated
in Euclidean space-time and thus cannot access quantities defined on the light-cone.
To circumvent this problem 
it has been proposed 
to calculate quasi parton distribution function (qPDF)
defined in terms of spatially separated quark bilinears \cite{Ji:2013dva}.
\begin{equation}
  \tilde q(x, P_z) = \frac{1}{4\pi}\int dz e^{-ixP^zz}
  \bra{H(P)}\bar\psi(z)\Gamma W_L(z, 0)\psi(0)\ket{H(P)},
\label{qpdf}
\end{equation}
where $\Gamma$ is either $\gamma_z$ or $\gamma_t$. For sufficiently
boosted hadron one can use Large 
Momentum Effective Theory (LaMET) \cite{Ji:2014gla} to relate qPDF to PDF:
\begin{equation}
\tilde q(x,\mu_L,P_z)=\int_{-1}^{+1}\frac{dy}{|y|}C\left(\frac{x}{y},\frac{P_z}{\mu},\frac{\mu_L}{P_z}\right)q(x,\mu).
\label{pdf2qpdf}
\end{equation}
Here $\mu_L$ and $\mu$ are the renormalization scales of the schemes in which qPDF and PDF are defined. For the later $\overline{MS}$ scheme
is used. The matching kernel has been calculated at 1-loop using cutoff scheme \cite{Xiong:2013bka} as well as in 
$\overline{MS}$ scheme \cite{Constantinou:2017sej,Liu:2018uuj,Stewart:2017tvs}.
For a comprehensive discussion on LaMET and related approaches see the recent review Ref. \cite{Cichy:2018mum}.
In this contribution we describe an exploratory study of pion PDF within LaMET framework.

\section{Lattice setup}
For calculations of PDF it is important to explore small values of $z$.
Therefore, we use lattices obtained using Highly Improved Staggered Quark (HISQ) action
with lattice spacing $0.06$ fm generated by HotQCD collaboration \cite{Bazavov:2014pvz}.
The lattice size is $48^3\times64$.
We use Wilson-Clover action for valence quarks on HYP 
smeared gauge configurations \cite{Hasenfratz:2001hp}
to avoid exceptional configurations. Very similar setup has been used by PDME
collaboration albeit for 2+1+1 flavor MILC configurations, see e.g. Ref. \cite{Liu:2018uuj}.
For the valence quarks we use quark masses corresponding to pion mass of about $300$ MeV.
For this quark mass we do no see exceptional configurations.
For the calculations of the two point and three point functions we 
we used All-Mode Averaging (AMA) \cite{Shintani:2014vja} with 32
sloppy calculations to one exact solve for each configuration.
For the sloppy inversion we use stopping criteria of $10^{-6}$. 
We performed calculations using $168$ gauge configurations for $z<0.48$ fm
and $52$ gauge configurations for $0.48~{\rm fm}<z<1$ fm.
In our study we neglect disconnected diagrams.

\section{Analysis of the two point function}
Obtaining a good signal for high momentum pion is non-trivial
as the noise becomes an issue at large time separations.
Therefore, the choice of the appropriate interpolating fields
is important. To increase overlap with the ground state we use Gaussian
sources for the pion. These are implemented either with Wuppertal smearing \cite{Gusken:1989ad}
or using Coulomb gauge. We find that 90 steps of Wuppertal smearing 
is the optimal choice that combines relatively fast approach of the effective
masses to a plateau with statistical errors that are not too large.
The source size corresponding to 90 steps of Wuppertal smearing is
about 0.3fm. The Coulomb gauge Gaussian sources of this size result in similar
errors for the effective masses. Since the use of Coulomb gauge Gaussian sources
turned out to be less expensive numerically we adopted this choice.

\begin{figure}
\includegraphics[scale=0.475]{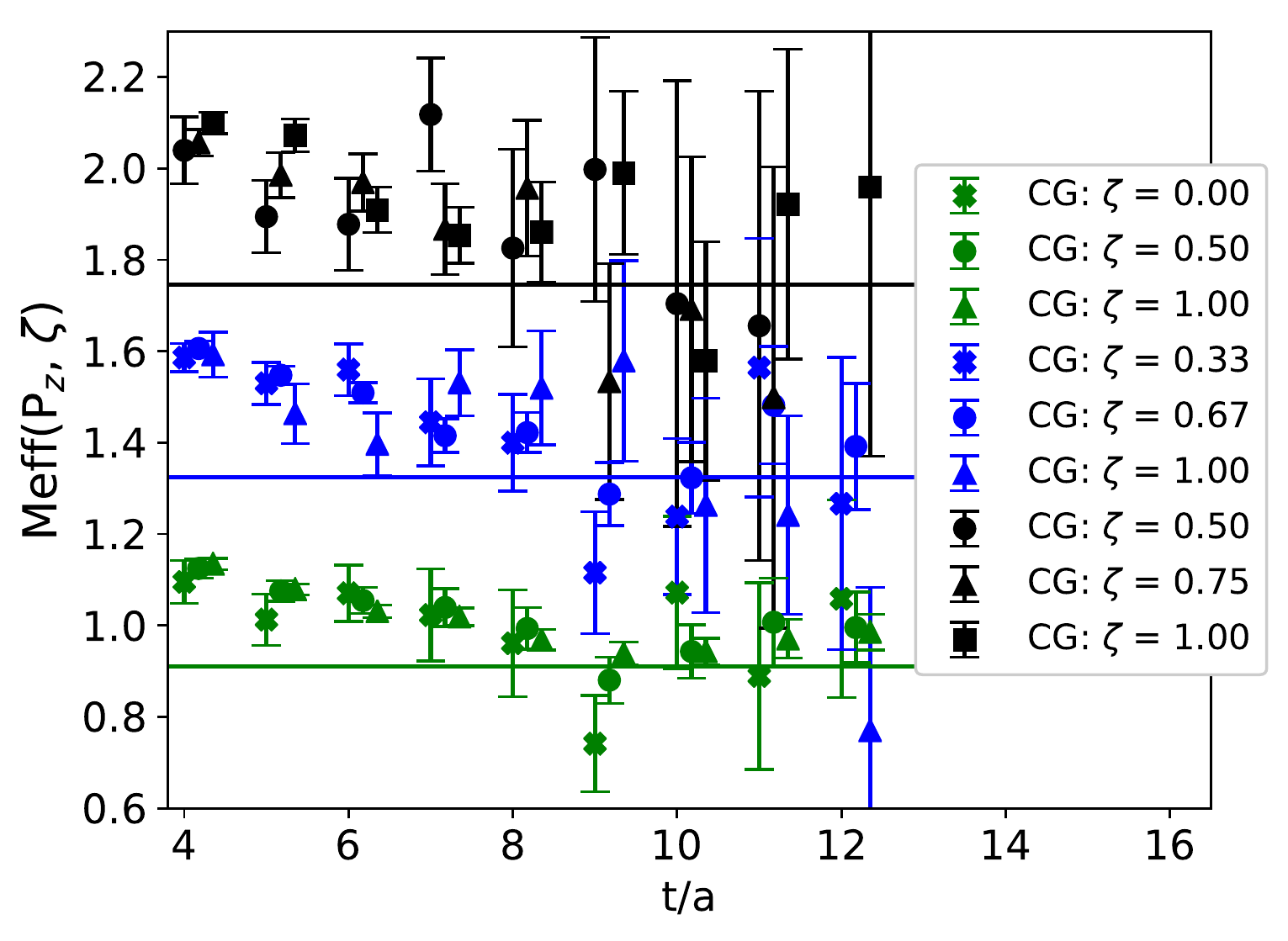}
\includegraphics[scale=0.475]{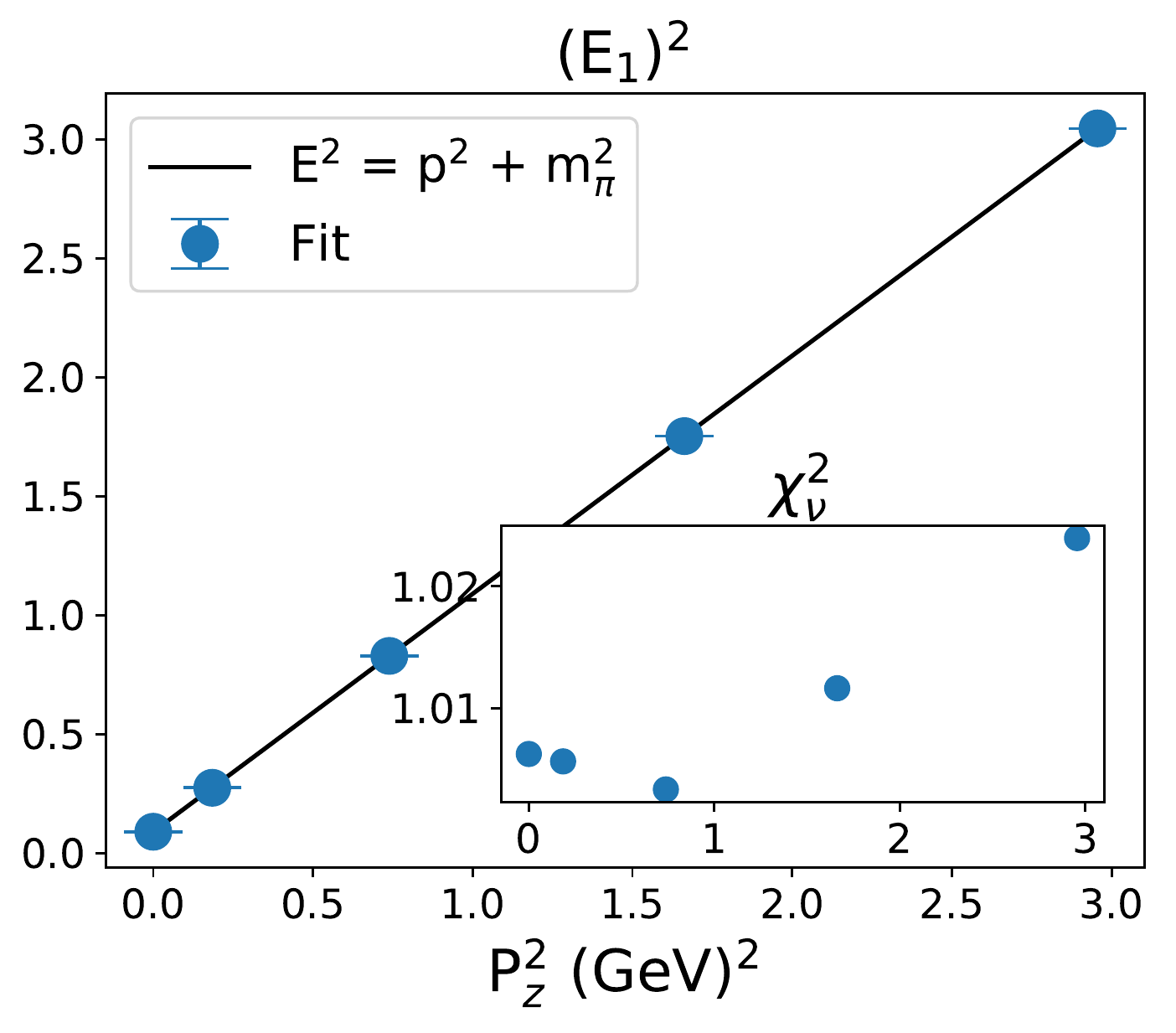}
\caption{Left: effective masses for different values of $\zeta$ with 50
configurations, Green, blue, and black points correspond to momentum 0.86,
1.29, and 1.72 GeV respectively. Right: the energy of the ground state as function
of $P_z$ from the two state fit.}
\label{fig:masses}
\end{figure}

For pion momenta of about $1$ GeV or larger the two point functions
is very noisy. To improve the signal following Ref. \cite{Bali:2016lva}
we use boosted sources, where 
the valence quarks are boosted to momentum
$\vec{k} = \zeta \vec{P}$, with $\vec{P}$ being the pion momentum and $\zeta$ is some number.
Naively one would expect that the optimal choice is $\zeta=0.5$.
In Figure \ref{fig:masses} we show the effective masses for
boosted Coulomb gauge Gaussian sources for different values of $\zeta$ at momentum 0.86
GeV, 1.29 GeV, and 1.72 GeV. At momentum 0.86 GeV we see significant improvement for
both $\zeta=0.5$ and $\zeta = 1.0$.
At 1.29 GeV the non-boosted sources are very noisy and are not shown
in the figure. Furthermore, $\zeta = 1.0$ turns out to be too large while $\zeta = 0.67$
yields the best results. At 1.72 GeV using $\zeta = 0.5$ is not sufficient, while 
the choices $\zeta = 0.75$ and $\zeta = 1.00$ give similar results. 
It is clear, however, that even with boosted sources extracting the ground state
at high momenta is difficult.
 
Next we performed two state fits for the pion correlation function
\begin{equation}
  C_{\text{2pt}}(P_z, t)=\sum_{i=1}^{2}2A_i e^{-\frac{1}{2}E_iT}\cosh{(E_i(T/2 - t))},
\end{equation}
to obtain the energies of the ground state and the excited state for different momenta $P_z$.
Here $T$ is the time extent of the lattice and $A_i = |\bra{i}\ket{\pi}|^2$.
The results for the ground state energy as function of $P_z$ are shown in Figure \ref{fig:masses}.
As one can see from the figure the determined energies follow the expected dispersion
relation. In Table 1 we present the difference of the excited state energy with
respect to the ground state energy 
for momenta 0 GeV, 0.86 GeV, 1.29 GeV, and 1.72 GeV.
Interestingly, we find that this energy gap is approximately independent of $P_z$.
\begin{table}
  \centering
  \caption{The energy difference between first excited state and the ground state for different $P_z$.}
  \begin{tabular}{| c | c | c | c | c | c |}
    \hline
    $P_z$(N$_\text{cfg}$)&0 GeV (52)&0.86 GeV (168)&1.29 GeV (168)&1.72 GeV (168) \\
    \hline
    $\Delta E_{2,1}$ & 1.39(38) GeV & 1.26(04) GeV & 1.15(08) GeV & 1.32(36) GeV \\
    \hline
  \end{tabular}
\end{table}

\section{Calculations of Three-Point Function}
To obtain the pion qPDF defined in Eq. (\ref{qpdf}) we consider the ratio
of the three-point to the two point function
\begin{equation}
  R(\Delta t, \tau, z; \Gamma) = \frac
  {\langle\pi(\vec{p},\Delta t)\mathcal{O}_{\Gamma}(z,\tau)\bar{\pi(0)}\rangle}
  {\langle\pi(\vec{p},\Delta t)\bar{\pi(0)}\rangle}
  = \frac{\sum_{n,n'}A_nA_{n'}^*e^{-E_n\Delta t}e^{-(E_{n'}-E_n)\tau}\bra{n}\mathcal{O}_{\Gamma}(z)\ket{n'}}{\sum_m |A_{m}|^2e^{-E_m \Delta t}}
\label{3pt}
\end{equation}
where $A_n = \bra{\pi}\ket{n}$,
\begin{equation}
\mathcal{O}_{\Gamma}(z) = \bar\psi(z)\Gamma W_L(z,0)\psi(0),
\label{defO}
\end{equation}
$\Delta t$ is the source-sink separation, and $\tau$ is the operator insertion
time, such that $0 < \tau < \Delta t$. 
For large $\Delta t$ and $\tau$ the above ratio gives the qPDF.
Inserting a complete set of states and truncating the sum to the two lowest terms
we write
\begin{equation}
  R(\Delta t, \tau, z; \Gamma) \sim
  \frac{\mathscr{M}(z) + \mathscr{A}(z)e^{-\Delta E_{2,1}\tau} +
    \mathscr{A}^\dagger(z)e^{-\Delta E_{2,1}(\Delta t-\tau)} +
    \mathscr{B}(z)e^{-\Delta E_{2,1}\Delta t} + ...}
  {1 + \mathscr{C}e^{-\Delta E_{2, 1}\Delta t} + ...}.
\label{3pt_large_t}
\end{equation}
Here, $\mathscr{M}(z) = \bra{1}\mathcal{O}_{\Gamma}(z)\ket{1}$ is the desired
quantity,
$\mathscr{A}(z) = \frac{A_1A_2^*}{|A_1|^2}\bra{1}\mathcal{O}_{\Gamma}(z)\ket{2}$,
$\mathscr{B}(z) = \mathscr{C}\bra{2}\mathcal{O}_{\Gamma}(z)\ket{2}$,
and $\mathscr{C} = \frac{|A_2|^2}{|A_1|^2}$.

In order to improve the signal we used one level of HYP smearing
in the Wilson line entering Eq. (\ref{defO}). The ratio $R(\Delta t, \tau, z; \Gamma)$
obtained with one level of HYP smearing is larger compared to the unsmeared
case. This is expected as smearing reduces the size of the self energy divergence in
the Wilson line, see e.g. Ref. \cite{Bazavov:2013zha}.
To obtain PDF one can use $\Gamma=\gamma_t$ or $\gamma_z$. The choice $\Gamma=\gamma_t$
has the advantage that in this case there is no mixing with the quark bilinear operator
with $\Gamma=1$ \cite{Constantinou:2017sej}. It also turns out that excited 
state contamination is smaller for $\Gamma=\gamma_t$. In what follows we discuss the 
calculations using one level of HYP smearing and $\Gamma=\gamma_t$.

\begin{figure}
\includegraphics[width=\textwidth]{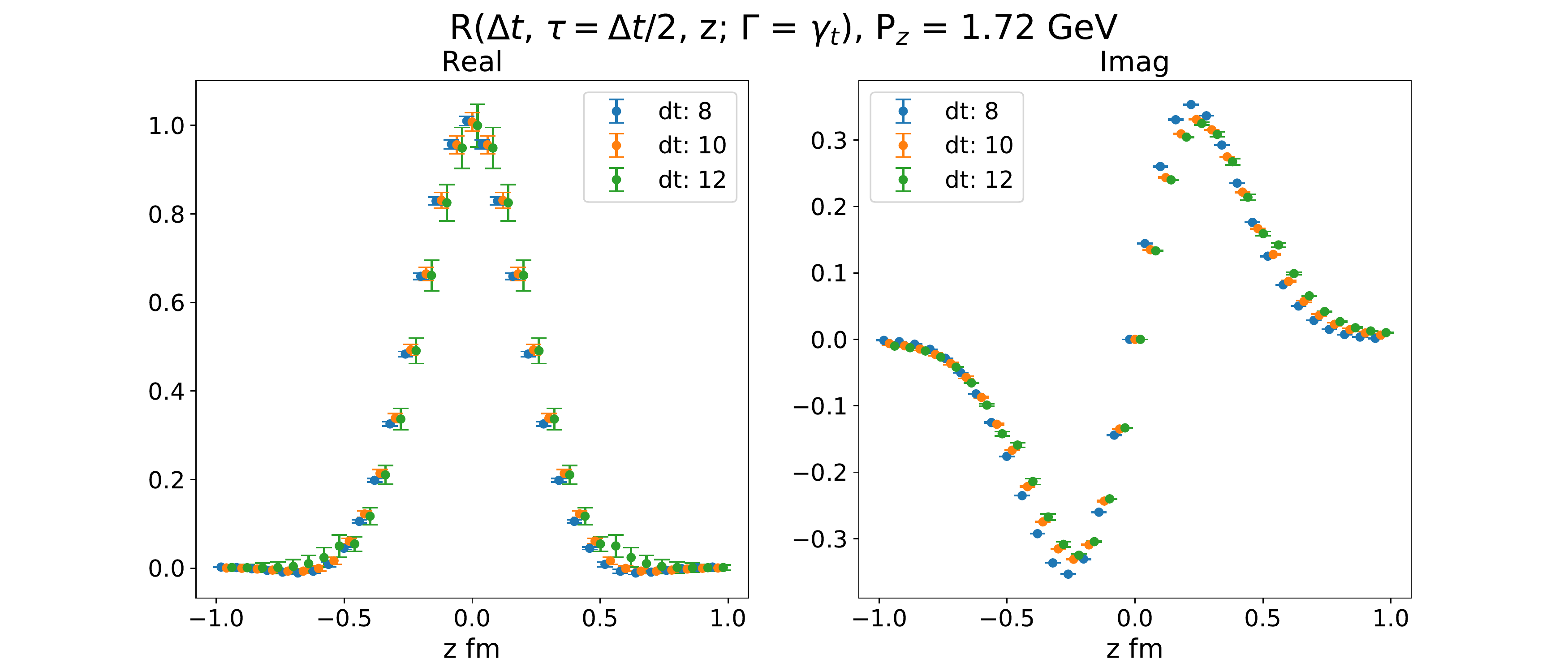}
\caption{$R(\Delta t, \tau, z; \gamma_t)$ at $P_z=1.72$ GeV as function of $z$ for $\Delta t/a=$ 8 (blue), 10 (orange),
    and 12 (green) and $\tau = \Delta t/2$.}
\label{fig:R_gammat}
\end{figure}

\begin{figure}
\includegraphics[scale=0.45]{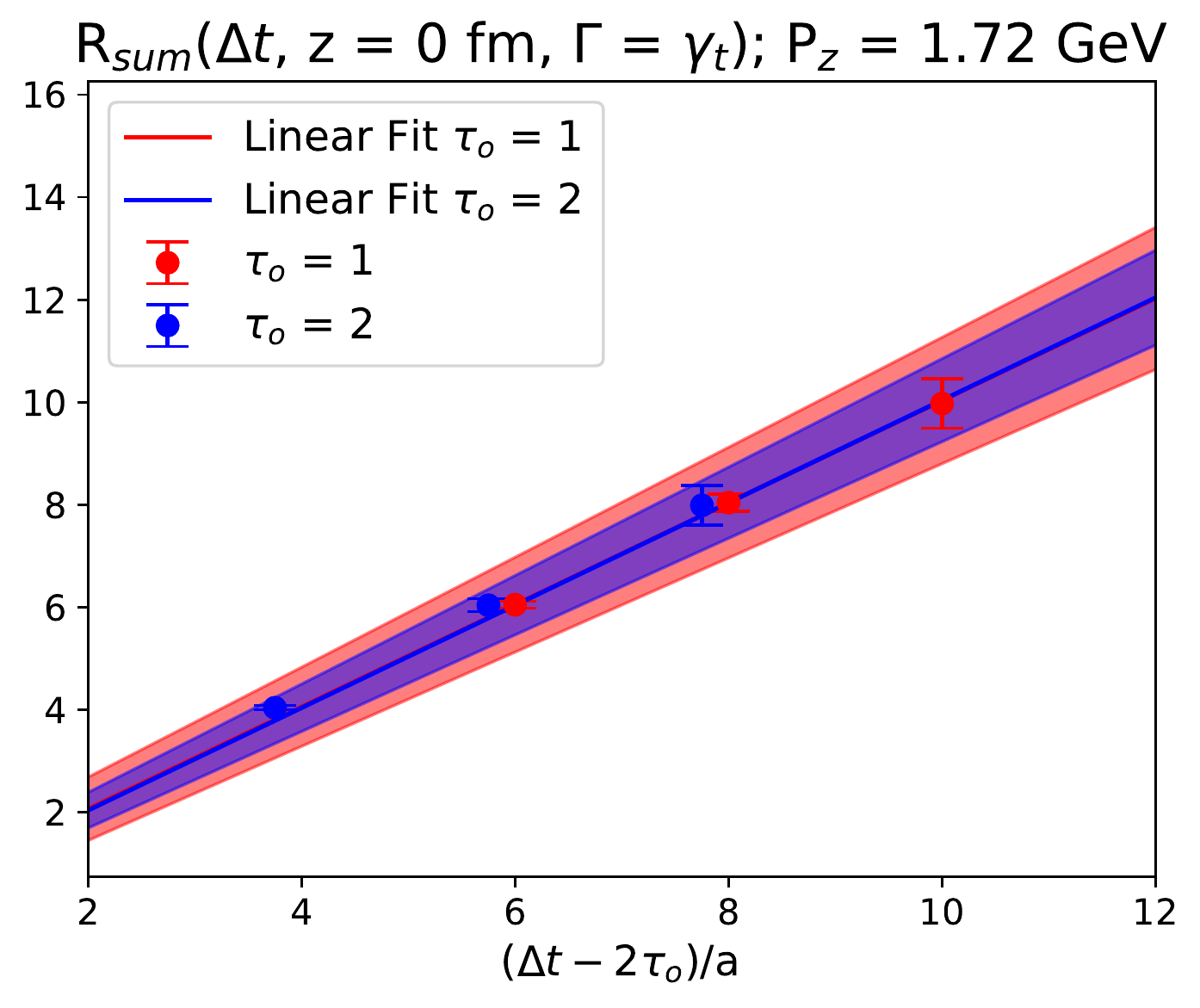}
\includegraphics[scale=0.45]{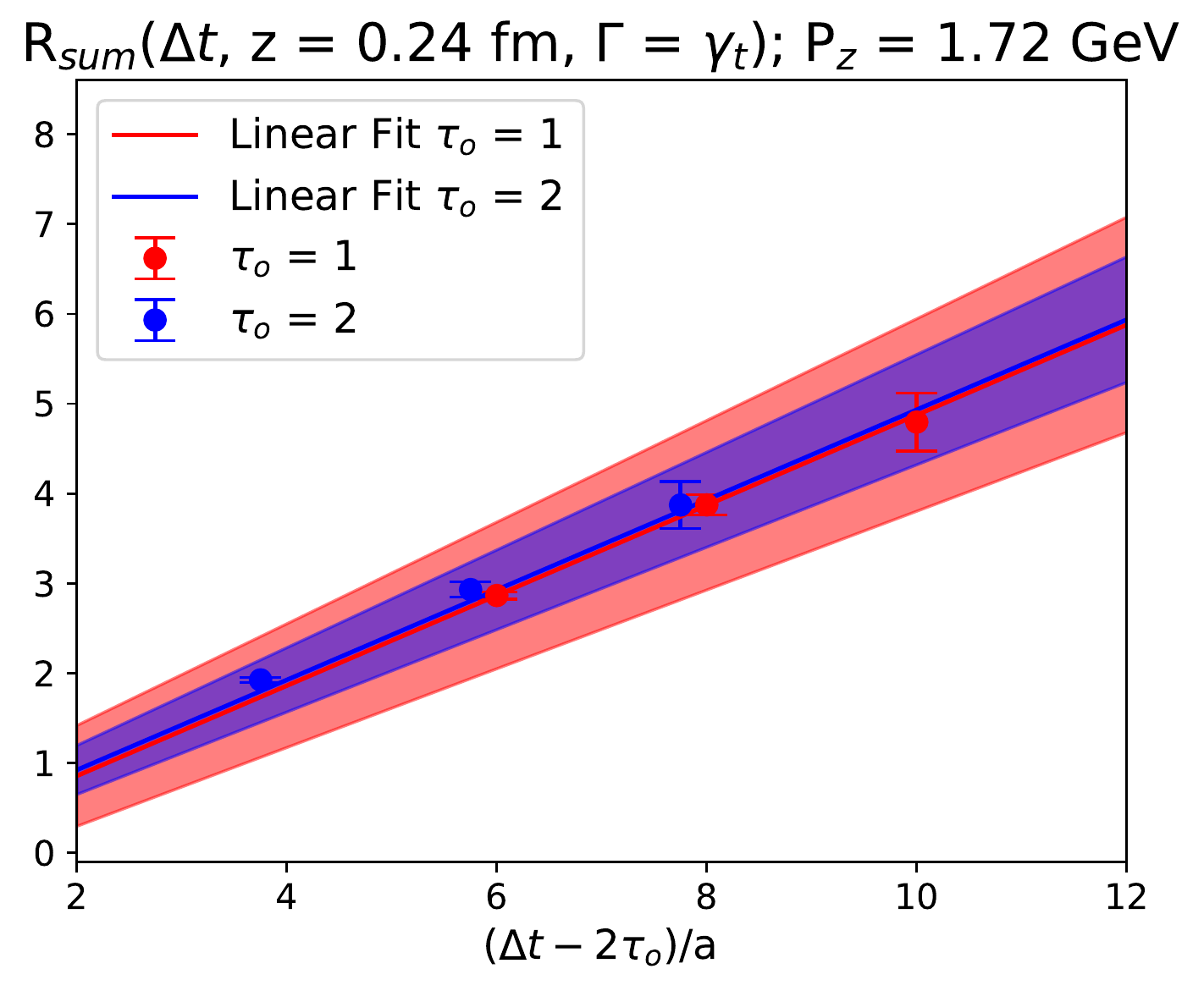}
\caption{$R_{\text{sum}}$ at $P_z=1.72$ GeV as function of $\Delta t - 2\tau_o$ for $z=0$ (left)
and $z = 0.24$ fm (right). The lines show the fit results and the bands show
the corresponding uncertainty.
Red and blue data points and bands are
      for results with $\tau_o$ = 1 and $\tau_o$ = 2 respectively.} 
\label{fig:sum}
\end{figure}

In Figure \ref{fig:R_gammat} we show the z dependence of $R(\Delta t,\tau,z;\gamma_t)$
for three source sink separations, $\Delta t=8,~10$ and $12$, and $\tau=\Delta t/2$.
The data points have been shifted horizontally for better visibility. 
We see a weak dependence on $\Delta t$ indicating that the contribution of the excited
states to $R(\Delta t,\tau,z;\gamma_t)$ is small.
To extract the
ground-state quasi-PDF matrix element we employ two fitting procedures used in
Refs. \cite{Abdel-Rehim:2015owa,Maiani:1987by}.
First we use  the summation method \cite{Maiani:1987by}.
Here one sums over all $\tau$ minus a certain number of
end points $\tau_o$ 
\begin{equation}
  R_{\text{sum}}(\Delta t, z; \Gamma) =
  \sum_{\tau=\tau_o}^{\Delta t - \tau_o}R(\Delta t, \tau, z; \Gamma) \sim
  (\mathscr{M} + \mathscr{B}e^{-\Delta E_{2,1} \Delta t})(\Delta t - 2\tau_o) + const.
\end{equation}
We calculate $R_{\text{sum}}(\Delta t, z; \Gamma)$ according to the above
equation and then perform a linear fit with respect to 
$\Delta t - 2\tau_o$. The slope obtained from the fit gives 
$\mathscr{M}$ for large enough $\Delta t$.
In Fig. \ref{fig:sum} we show the results of the summation method for 
$z = 0$ fm and $z = 0.24$ fm and $P_z=1.72$ GeV using $\tau_o = 1$ and $\tau_o = 2$.
For both values of $z$ the choice $\tau_o = 2$ gives the most precise
result. 
The second method relies on simultaneous fit of
the $\Delta t$ and $\tau$ dependence of the ratio $R(\Delta t,\tau,z;\gamma_t)$
to the form given by  Eq. (\ref{3pt_large_t}),
which we refer to as the two-state fit \cite{Abdel-Rehim:2015owa}. 
Here we use
$\Delta E_{2,1}$ obtained from the two-point correlator and summarized in Table 1
and treat $\mathscr{M}$, $\mathscr{A}$, and $\mathscr{B}$ as fit parameters.

\begin{figure}
\includegraphics[width=\textwidth]{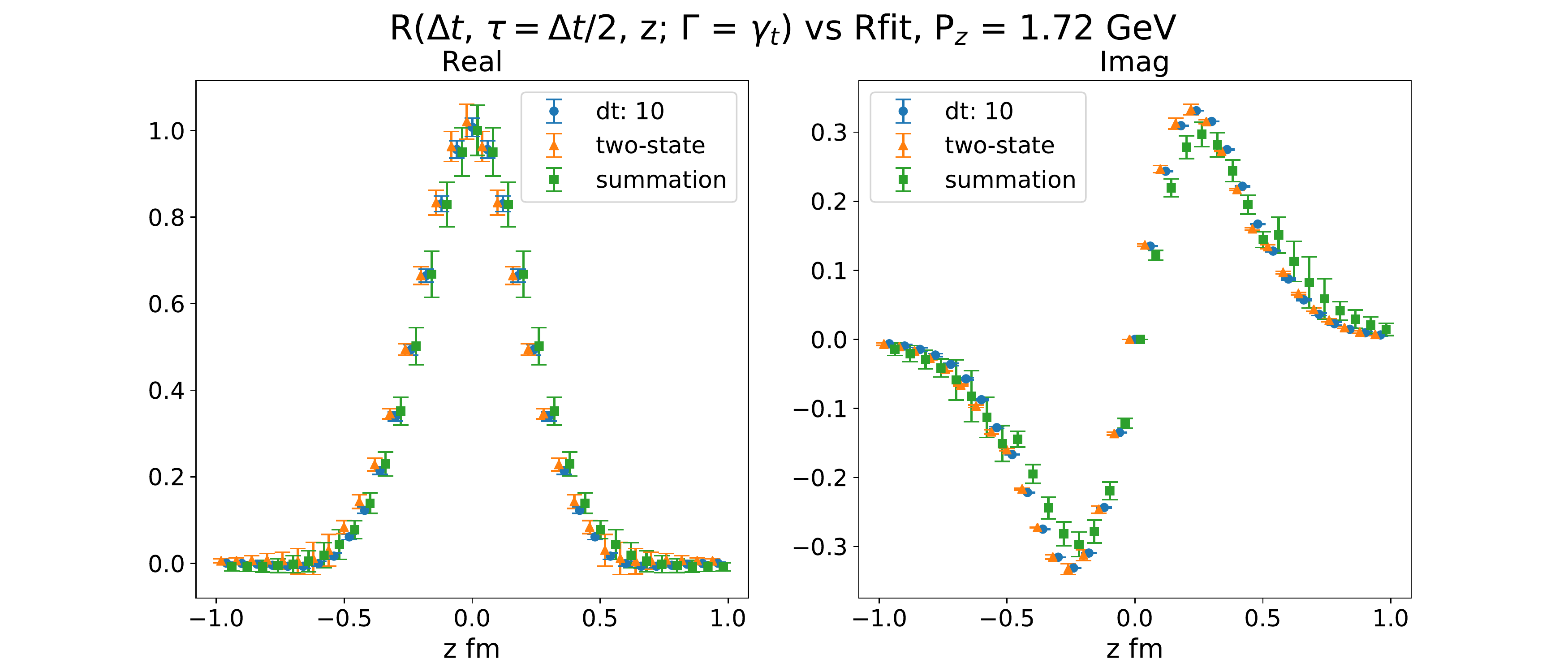}
\caption{
The real part (left) and the imaginary  part (right) of $\mathscr{M}$ for $P_z=1.72$ GeV.
obtained using the summation method and the two-state fit. We also
show our results for $R(\Delta t/a=10,\tau/a=5,z;\gamma_t)$.}
\label{fig:M}
\end{figure}

In Fig. \ref{fig:M} we show our results for $\mathscr{M}$ obtained using the summation
method and two-state fit. We also compare $\mathscr{M}$ 
with the $R(\Delta t,\tau,z;\gamma_t)$ evaluated at $\Delta t$ = 10 with the $\tau = \Delta t/2$. 
For the real part the two methods extracting $\mathscr{M}$ agree within errors and also agree
with $R(\Delta t=10,\tau=5,z;\gamma_t)$. This means that excited state contributions
are under control.
The imaginary part of $\mathscr{M}$ obtained with 
the summation method does differ somewhat to the results with
two-state fit and ${\rm Im} R(\Delta t/a=10,\tau/a=5,z;\gamma_t)$, 
meaning that excited states have some effects in the imaginary part. 

\section{Calculating the parton distribution}

\begin{figure}
\includegraphics[scale=0.7]{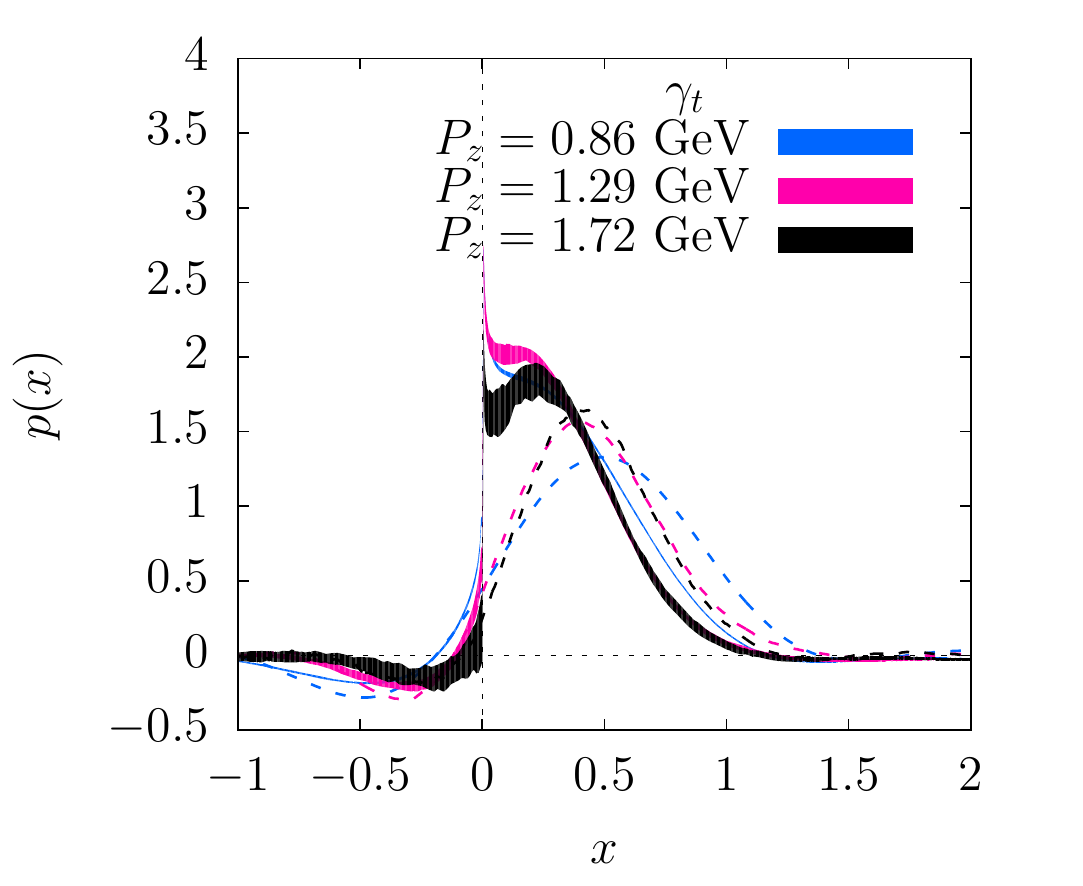}
\includegraphics[scale=0.7]{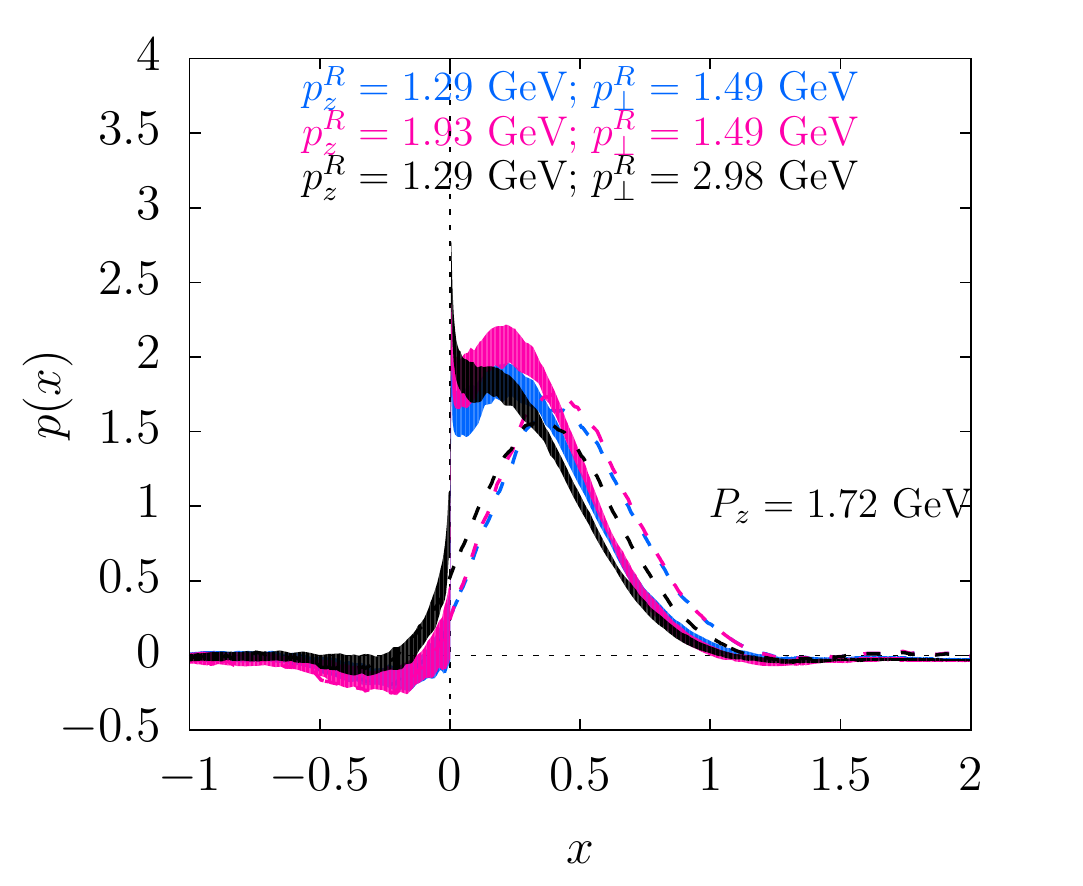}
\caption{Left: PDF and qPDF for different values of $P_z$. Right: the dependence
of PDF and qPDF on the RI-MOM renormalizations scales for $P_z=1.72$ GeV.}
\label{fig:pdf}
\end{figure}

The matrix element $\mathscr{M}$ that defines the qPDF needs to be renormalized.
We perform the renormalization using regularization independent momentum subtraction
(RI-MOM) scheme. To define the renormalization constant one calculates the
expectation value of the non-local quark bilinear in Eq. (\ref{defO}) on off-shell
quark states: $\Lambda(p,z)=\bra{p} O_{\gamma_t}(z) \ket{p}$. The renormalization
condition is defined such that the renormalized matrix element
$\Lambda^{R}(p,z)=Z(z,p_z^R,p^R) \Lambda(p,z)$ satisfies the condition
${\rm Tr} \slashed p \Lambda^{R}(p,z)|_{p=p^R}=12 p_t^R exp(-i p_z^R z)$, i.e.
it equals to the tree level result for $p=p_R$. The RI-MOM scheme here depends
on two renormalization scales: $p_z^R$ and $p_R^2=(p_z^R)^2+(p_{\perp}^R)^2$, because the $z$ direction plays
a special role. In other words, for RI-MOM scheme $\mu_L=\{p_z^R,p_R^2\}$.
A more detailed discussion of our RI-MOM renormalization procedure is
given in Ref. \cite{Karthik:2018wmj}.
Multiplying  the bare matrix element by $Z(z,p_z^R,p_R^2)$ we get the renormalized
matrix element $\mathscr{M}^R$, which then can be used to calculate the qPDF according to Eq. (\ref{qpdf}).
In our preliminary study we used $R(\Delta t/a=10,\tau/a=5,z;\gamma_t)$ as proxy for $\mathscr{M}$.
As discussed in the previous section the excited state contamination is small for $\Delta t/a=10$. 
To perform the Fourier transformation in Eq. (\ref{qpdf}) we need to information about $\mathscr{M}^R$ for all $z$.
However, our numerical calculations only cover the value of $|z|$ up to $1$ fm. Since $\mathscr{M}^R$
decays rapidly at large $z$ we assume that it vanishes for $|z/a|=20$ and perform interpolation
of the lattice results on $\mathscr{M}^R$ with this constraint. Using this interpolation we 
calculate the qPDF $\tilde q(x,P_z,p_z^R,p_R)$  \cite{Karthik:2018wmj}. The resulting qPDF
are shown in Fig. \ref{fig:pdf} as dashed lines. We checked that the numerical results
do not change much if we assume that the coordinate space qPDF vanishes at $|z/a|=24$.
To calculate PDF from qPDF we invert Eq. (\ref{pdf2qpdf}) to leading order in $\alpha_s$.
The matching kernel $C$ entering Eq. (\ref{pdf2qpdf}) for $\Gamma=\gamma_t$ and qPDF in RI-MOM
scheme has been calculated at 1-loop in Ref. \cite{Liu:2018uuj}, and we make use of the corresponding result
in our analysis. Our results for pion PDF are shown in Fig. \ref{fig:pdf}. We see that
for $P_z \ge 1.29$ GeV the dependence of the reconstructed PDF on $P_z$ is very small.
The PDF should be independent on RI-MOM scale parameters $p_z^R$ and $p_R^2$ because this
dependence cancels out between qPDF and the matching kernel $C$. In practice, however,
this cancellation is not exact as the matching kernel is only know at 1-loop. From Fig. \ref{fig:pdf}
we see that the dependence on $p_z^R$ and $p^R$ is rather mild, which is encouraging and 
indicates that the outlined strategy for calculating PDF is viable.

\section{Conclusions}
In this contribution we presented preliminary calculations of quark distribution inside the pion
from lattice QCD based on Large Momentum Effective Theory approach by Ji.
We used fine lattices ($a=0.06$fm) in order to utilize  the pertubative matching between qPDF and PDF.
To obtain the renormalized qPDF we used the non-perturbative RI-MOM scheme. Obtaining the ground state
signal for the fast moving pion is very challenging and in order to reach this goal we used
momentum boosted sources. With these we were able to reach pion momenta up to $1.72$ GeV.
In order to perform calculations at even larger values of $P_z$ significantly more statistics
will be needed.

\acknowledgments
This work was supported by the U.S. Department of Energy
under contract No. DE-SC0012704, BNL LDRD project No. 16-37 and Scientific Discovery
through Advance Computing (SCiDAC) award ”Computing the Properties of Matter with
Leadership Computing Resources”. The computations were carried out using USQCD
facilities at JLab and BNL under a USQCD type-A project. This research also used an
award of computer time provided by the INCITE program at the Oak Ridge Leadership
Computing Facility, which is a DOE Office of Science User Facility supported under
Contract DE-AC05-00OR22725.

\bibliographystyle{JHEP}
\bibliography{ref.bib}

\providecommand{\href}[2]{#2}\begingroup\raggedright\begin{thebibliography}{10}

\bibitem{Ji:2013dva}
X.~Ji, \emph{{Parton Physics on a Euclidean Lattice}},
  \href{https://doi.org/10.1103/PhysRevLett.110.262002}{\emph{Phys. Rev. Lett.}
  {\bfseries 110} (2013) 262002}
  [\href{https://arxiv.org/abs/1305.1539}{{\ttfamily 1305.1539}}].

\bibitem{Ji:2014gla}
X.~Ji, \emph{{Parton Physics from Large-Momentum Effective Field Theory}},
  \href{https://doi.org/10.1007/s11433-014-5492-3}{\emph{Sci. China Phys. Mech.
  Astron.} {\bfseries 57} (2014) 1407}
  [\href{https://arxiv.org/abs/1404.6680}{{\ttfamily 1404.6680}}].

\bibitem{Xiong:2013bka}
X.~Xiong, X.~Ji, J.-H. Zhang and Y.~Zhao, \emph{{One-loop matching for parton
  distributions: Nonsinglet case}},
  \href{https://doi.org/10.1103/PhysRevD.90.014051}{\emph{Phys. Rev.}
  {\bfseries D90} (2014) 014051}
  [\href{https://arxiv.org/abs/1310.7471}{{\ttfamily 1310.7471}}].

\bibitem{Constantinou:2017sej}
M.~Constantinou and H.~Panagopoulos, \emph{{Perturbative renormalization of
  quasi-parton distribution functions}},
  \href{https://doi.org/10.1103/PhysRevD.96.054506}{\emph{Phys. Rev.}
  {\bfseries D96} (2017) 054506}
  [\href{https://arxiv.org/abs/1705.11193}{{\ttfamily 1705.11193}}].

\bibitem{Liu:2018uuj}
Y.-S. Liu, J.-W. Chen, L.~Jin, H.-W. Lin, Y.-B. Yang, J.-H. Zhang et~al.,
  \emph{{Unpolarized quark distribution from lattice QCD: A systematic analysis
  of renormalization and matching}},
  \href{https://arxiv.org/abs/1807.06566}{{\ttfamily 1807.06566}}.

\bibitem{Stewart:2017tvs}
I.~W. Stewart and Y.~Zhao, \emph{{Matching the quasiparton distribution in a
  momentum subtraction scheme}},
  \href{https://doi.org/10.1103/PhysRevD.97.054512}{\emph{Phys. Rev.}
  {\bfseries D97} (2018) 054512}
  [\href{https://arxiv.org/abs/1709.04933}{{\ttfamily 1709.04933}}].

\bibitem{Cichy:2018mum}
K.~Cichy and M.~Constantinou, \emph{{A guide to light-cone PDFs from Lattice
  QCD: an overview of approaches, techniques and results}},
  \href{https://arxiv.org/abs/1811.07248}{{\ttfamily 1811.07248}}.

\bibitem{Bazavov:2014pvz}
{\scshape HotQCD} collaboration, A.~Bazavov et~al., \emph{{Equation of state in
  ( 2+1 )-flavor QCD}},
  \href{https://doi.org/10.1103/PhysRevD.90.094503}{\emph{Phys. Rev.}
  {\bfseries D90} (2014) 094503}
  [\href{https://arxiv.org/abs/1407.6387}{{\ttfamily 1407.6387}}].

\bibitem{Hasenfratz:2001hp}
A.~Hasenfratz and F.~Knechtli, \emph{{Flavor symmetry and the static potential
  with hypercubic blocking}},
  \href{https://doi.org/10.1103/PhysRevD.64.034504}{\emph{Phys. Rev.}
  {\bfseries D64} (2001) 034504}
  [\href{https://arxiv.org/abs/hep-lat/0103029}{{\ttfamily hep-lat/0103029}}].

\bibitem{Shintani:2014vja}
E.~Shintani, R.~Arthur, T.~Blum, T.~Izubuchi, C.~Jung and C.~Lehner,
  \emph{{Covariant approximation averaging}},
  \href{https://doi.org/10.1103/PhysRevD.91.114511}{\emph{Phys. Rev.}
  {\bfseries D91} (2015) 114511}
  [\href{https://arxiv.org/abs/1402.0244}{{\ttfamily 1402.0244}}].

\bibitem{Gusken:1989ad}
S.~Gusken, U.~Low, K.~H. Mutter, R.~Sommer, A.~Patel and K.~Schilling,
  \emph{{Nonsinglet Axial Vector Couplings of the Baryon Octet in Lattice
  {QCD}}}, \href{https://doi.org/10.1016/S0370-2693(89)80034-6}{\emph{Phys.
  Lett.} {\bfseries B227} (1989) 266}.

\bibitem{Bali:2016lva}
G.~S. Bali, B.~Lang, B.~U. Musch and A.~Schäfer, \emph{{Novel quark smearing
  for hadrons with high momenta in lattice QCD}},
  \href{https://doi.org/10.1103/PhysRevD.93.094515}{\emph{Phys. Rev.}
  {\bfseries D93} (2016) 094515}
  [\href{https://arxiv.org/abs/1602.05525}{{\ttfamily 1602.05525}}].

\bibitem{Bazavov:2013zha}
A.~Bazavov and P.~Petreczky, \emph{{Static meson correlators in 2+1 flavor QCD
  at non-zero temperature}},
  \href{https://doi.org/10.1140/epja/i2013-13085-8}{\emph{Eur. Phys. J.}
  {\bfseries A49} (2013) 85} [\href{https://arxiv.org/abs/1303.5500}{{\ttfamily
  1303.5500}}].

\bibitem{Abdel-Rehim:2015owa}
A.~Abdel-Rehim et~al., \emph{{Nucleon and pion structure with lattice QCD
  simulations at physical value of the pion mass}},
  \href{https://doi.org/10.1103/PhysRevD.92.114513,
  10.1103/PhysRevD.93.039904}{\emph{Phys. Rev.} {\bfseries D92} (2015) 114513}
  [\href{https://arxiv.org/abs/1507.04936}{{\ttfamily 1507.04936}}].

\bibitem{Maiani:1987by}
L.~Maiani, G.~Martinelli, M.~L. Paciello and B.~Taglienti, \emph{{Scalar
  Densities and Baryon Mass Differences in Lattice {QCD} With Wilson
  Fermions}}, \href{https://doi.org/10.1016/0550-3213(87)90078-2}{\emph{Nucl.
  Phys.} {\bfseries B293} (1987) 420}.

\bibitem{Karthik:2018wmj}
N.~Karthik, T.~Izubichi, L.~Jin, C.~Kallidonis, S.~Mukherjee, P.~Petreczky
  et~al., \emph{{Renormalized quasi parton distribution function of pion}},
  {\emph{PoS} {\bfseries LATTICE2018} (2018) 109}
  [\href{https://arxiv.org/abs/1811.06075}{{\ttfamily 1811.06075}}].

\end{thebibliography}\endgroup

\end{document}